\documentstyle [twocolumn,aps,prbbib]{revtex}
\include{psfig}
\begin{document}
\draft
\title{
Unified Treatment of Asymptotic van der Waals Forces
}
\author{Erika Hult, Henrik Rydberg, and Bengt I. Lundqvist}
\address{Department of Applied Physics,  
Chalmers University of Technology and G\"{o}teborg University, 
S-412 96 G\"{o}teborg, Sweden
}
\author{David C. Langreth}
\address{
Department of Physics and Astronomy,
Rutgers University,
Piscataway, New Jersey 08854-8019
}
\maketitle
\begin{abstract}
In a framework for long-range density-functional theory
we present a unified full-field treatment of the asymptotic van der Waals
interaction for atoms, molecules, surfaces, and other objects. The only 
input needed consists of the electron densities of the interacting
fragments and 
the static polarizability or the static image plane, which can be easily 
evaluated in a ground-state density-functional calculation for each
fragment. Results for 
separated atoms, molecules, and 
for atoms/molecules outside surfaces
are in agreement with
those of other, more elaborate, calculations. 
\end{abstract}


The ubiquitous van der Waals interaction needs an efficient and accurate 
description in many contexts such as interacting noble-gas atoms, van 
der Waals complexes, physisorption, interacting macroscopic neutral bodies, 
liquid-crystal interactions, solute-solvent interactions, and soft-condensed
matter. For dense matter 
the density-functional theory (DFT),~\cite{HoKo64} with its 
local-density (LDA)~\cite{KoSh65,GuLu76}
and generalized-gradient approximations 
(GGA),~\cite{LaMe81,Becke88,Peetal92,PeBuEr96}
is a clear success.
Ground-state and thermodynamic properties of increasingly more complex 
systems are now being calculated with a practically very useful accuracy.
As the 
world around contains far more objects than just hard solids, 
a generalization of these methods to also account for the van der Waals 
forces is in great demand. These forces are an inherent property of the exact 
DFT~\cite{LuAnShChLa95} and it is thus a question of providing an
approximate van der Waals
density functional that is generally
applicable, efficient, and accurate, and that per definition is a functional 
of the density only. Earlier we~\cite{AnLaLu96,HuAnLuLa96} and
others~\cite{DoDi96,KoMeMa97} have proposed such
functionals and shown them to give useful results in significant applications.
For a review, see Ref.~\onlinecite{Australia}. Until now, however,
there has been a
certain  asymmetry in our treatment of ``small'' and ``large'' objects,
respectively, which will be remedied in this paper.

The starting point for our functionals is the exact
expression for the exchange-correlation energy $E_{\rm xc}$ as
an integral over the coupling constant
$\lambda e^2$, the so-called {\em adiabatic connection formula}
(ACF)~\cite{GuLu76,LaPe7577}
\begin{eqnarray}
   E_{\rm xc}[n]&=&\frac{1}{2} \int d^3r d^3r^{\prime} \, \frac{e^2}
                {| {\bf r}- {\bf r}^{\prime}|} \times \nonumber \\
            & & \int_0^1 d \lambda \,
                \left[ \langle \tilde{n}({\bf r}) \tilde{n}({\bf
                r}^{\prime}) \rangle_{n, \lambda} - \delta
                ({\bf r}-{\bf r}^{\prime}) \langle n({\bf r})
                \rangle \right]\,,
\label{eq:ACF}
\end{eqnarray}
where $\tilde{n}=\hat{n}-n$, $\hat{n}$ being the density operator,
and $\langle \dots \rangle_{n, \lambda}$ means that the integration
is performed with a potential $V_{\lambda}$ present, keeping the
density equal to $n({\bf r})$.
To second
order perturbation theory in the interaction $V_{ab}$ between two separated 
objects $a$ and $b$ the ACF can be cast into the
form~\cite{LuAnShChLa95,Dobson94}
\begin{eqnarray}
\label{eq:vdW1}
\nonumber
        \Delta E_{\rm xc}({\bf R}) & =& E_{\rm xc}^{\infty} -\int\!
        \int\! \int\!
                \int\! d^{3}r_1 d^{3}r_2  d^{3}r_3 d^{3}r_4 \, \times
         \nonumber \\
        & &      V_{ab}({\bf R}+{\bf r}_1-{\bf r}_2) V_{ab}({\bf R}+{\bf r}_3 -
                {\bf r}_4)\times \nonumber \\
        & & \int_{0}^{\infty}\! \frac{du}{2\pi} \,
                \Pi_{a}({\bf r}_1,{\bf r}_3;iu)
                \Pi_{b}({\bf r}_2, {\bf r}_4;iu)\, .
\end{eqnarray} 
Our evaluation of this energy is based on two approximations.
First we introduce a local dielectric function that
depends on the local electron density.
Second, we limit the volumes of the interacting objects by using
a cutoff, an idea first introduced by Rapcewicz and
Ashcroft,~\cite{RaAs91} outside which the response to an electric
field is defined to be zero.
We thus have the dielectric function
\begin{equation}
\epsilon(\omega; n({\bf r}))=1-\kappa(n({\bf r})) 
                             \frac{\omega_p^2(n({\bf r}))}{\omega^2}\,,
\label{eq:epsdef}
\end{equation}
where
\begin{equation}
\omega_p^2(n({\bf r})) = 4 \pi e^2 n({\bf r})/m \,.
\label{omega_p}
\end{equation}
The cutoff is implemented via the function $\kappa(n(r))$, which is either
unity or zero, following the notion discussed
earlier~\cite{AnLaLu96}
that the local approximation for dielectric response greatly
exaggerates the response in the low-density tails, where
it is better to assume no response at all.

These approximations are common for all of our three model systems --
interacting atoms or molecules,~\cite{AnLaLu96,AnRy97} an atom 
or molecule interacting with a planar surface,~\cite{HuAnLuLa96,HuKi97}
and finally two 
interacting surfaces.~\cite{AnHuApLaLu98}
However, we have earlier treated the electrodynamics on 
different levels for ``small'' and ``large'' objects. 
Normally, local electrodynamics means a local relationship between
the polarization ${\bf P}$ and the total electric field ${\bf E}$
\begin{equation}
 {\bf P}({\bf r},\omega) = \frac{1}{4 \pi} [\epsilon(\omega;n({\bf r})) -1]
 \,{\bf E}({\bf r},\omega)\,,
\label{eq:PfromE}
\end{equation}
which we use for surfaces.
For atoms, however, the calculations get somewhat more complicated
than in the surface case when using Eq.~(\ref{eq:PfromE}). For instance, the
electrodynamics must be solved numerically for each frequency.
Our earlier approach for atoms and molecules has approximated
the local polarization by
\begin{equation}
{\bf P}({\bf r},\omega) = \frac{1}{4 \pi} [\epsilon(\omega; n({\bf r})) -1]
 \frac{{\bf E}_{\rm applied}({\bf r},\omega)}{\epsilon(\omega; n({\bf r}))}\,,
\label{eq:PfromEappl}
\end{equation}
which implies the relation 
\begin{equation}
 {\bf E}({\bf r},\omega)= {\bf E}_{\rm applied}({\bf r},\omega)/
                          \epsilon(\omega; n({\bf r})) \,,
\end{equation}
which is certainly wrong for macroscopic objects,
but gives surprisingly good results for atoms and
molecules.~\cite{AnLaLu96,AnRy97}
However, the optimum cutoff position defined by $\kappa(n(r))$ is
somewhat larger for the approximate electrodynamics, and this fact
seems to mitigate some of the deficiencies of this previous approximation.
To obtain a unified treatment for different objects, and also to test our
approximation for the dielectric function, this paper presents
the electrodynamics using Eq.~(\ref{eq:PfromE}) also for atoms
and molecules.
We apply it to the asymptotic van der Waals interaction of separated
atoms, molecules, and parallel surfaces,~\cite{AnHuApLaLu98}
and show the results to
be in agreement with those of other, more elaborate, calculations.

For two widely separated atoms or molecules the van der Waals energy
is given by $E_{\rm vdW}=-C_6/R^6$, where the van der Waals coefficient
is~\cite{MaSt62,MaKe69}
\begin{equation}
   C_6 = \frac{3}{\pi} \int_0^{\infty} \! du \, \alpha_1(iu) \alpha_2(iu) \,
\label{eq:C6}
\end{equation}
and $\alpha_j(iu)$ is the polarizability of atom $j$. 
To calculate $\alpha(iu)$ we solve $\nabla \cdot {\bf D}
({\bf r})=0$, for each frequency in the presence of an applied
constant electric field ${\bf E}_0$.
The displacement ${\bf D}({\bf r})$ is given by
$\epsilon(\omega;n({\bf r})) {\bf E}
({\bf r})$ and ${\bf E}({\bf r})=-\nabla \phi ({\bf r})$. Thus we solve the
equation
\begin{equation}
  \nabla \cdot (\epsilon(\omega;n({\bf r})) \nabla \phi ({\bf r})) = 0 \,,
\label{eq:div}
\end{equation}
where $\epsilon(\omega;n({\bf r}))$ is given by Eqs.~(\ref{eq:epsdef}) and
(\ref{omega_p}), and with the boundary condition that ${\bf E}$ approaches
${\bf E}_0$ at infinity.

The cutoff function $\kappa(n({\bf r}))$ must however first be defined.
For surfaces the cutoff is found by taking the static limit
of the surface response,~\cite{HuAnLuLa96} which implies
that the cutoff should be defined
by the static image plane $d(0)$. 
In order to have a common cutoff scheme for both atoms, molecules,
and surfaces,
and in addition to implement the requirement introduced for surfaces
that the static polarization response be accurate, it is expedient to
simplify the scheme used previously for atoms and molecules. 
This is done by choosing the
cutoff function $\kappa$ according to 
\begin{equation}
  \kappa({\bf r})=\theta(n({\bf r})-c) \,,
\label{cutoffimplement}
\end{equation}
where $c$ is a constant. 

For atoms and molecules, compared with the original scheme that
uses both the density and its gradient,
the practical effect of (\ref{cutoffimplement})
is to eliminate any cutoff in the intra-shell regions. We have found that
inclusion of the cutoff in the intra-shell regions as before results
in a median reduction of the predicted values of the $C_6$ coefficients
of $12\%$ for the atom pairs calculated here. 
The extent to which these intra-shell corrections should be included
even in principle is arguable, and since they are small, 
we have therefore opted for the
simpler scheme (\ref{cutoffimplement}). Adapting the analogue of
the procedure used when the full-field scheme is applied to
surfaces,~\cite{HuAnLuLa96} we fix the constant $c$ in (\ref{cutoffimplement})
so that the static polarizibilities are accurate. For a spherically
symmetric species, this means that the volume $V$ inside which the
step function in (\ref{cutoffimplement}) is nonvanishing is
simply $V=(4\pi/3)\alpha(0)$. For species without spherical symmetry,
we choose $c$ so that the isotropic polarizabilities, $\bar{\alpha}(0)$
[see Eq.~(\ref{alphabar})],
are correct. This scheme seems to underestimate the anisotropy
of the molecular polarizability; if an accurate anisotropy is important
the cutoff function should be modified so that the elements of the
diagonalized static polarizability tensor are reproduced.

\begin{figure}
\caption{Calculated van der Waals coefficients $C_6$ (Ry atomic units)
for all possible pairs of the atoms in Table~\protect\ref{sametab} plotted
against corresponding values from other 
calculations~\protect\cite{BiPi93,StCe85,MaKu79,KoHa91,TaNoCe76}.\\}
\centerline{\psfig{figure=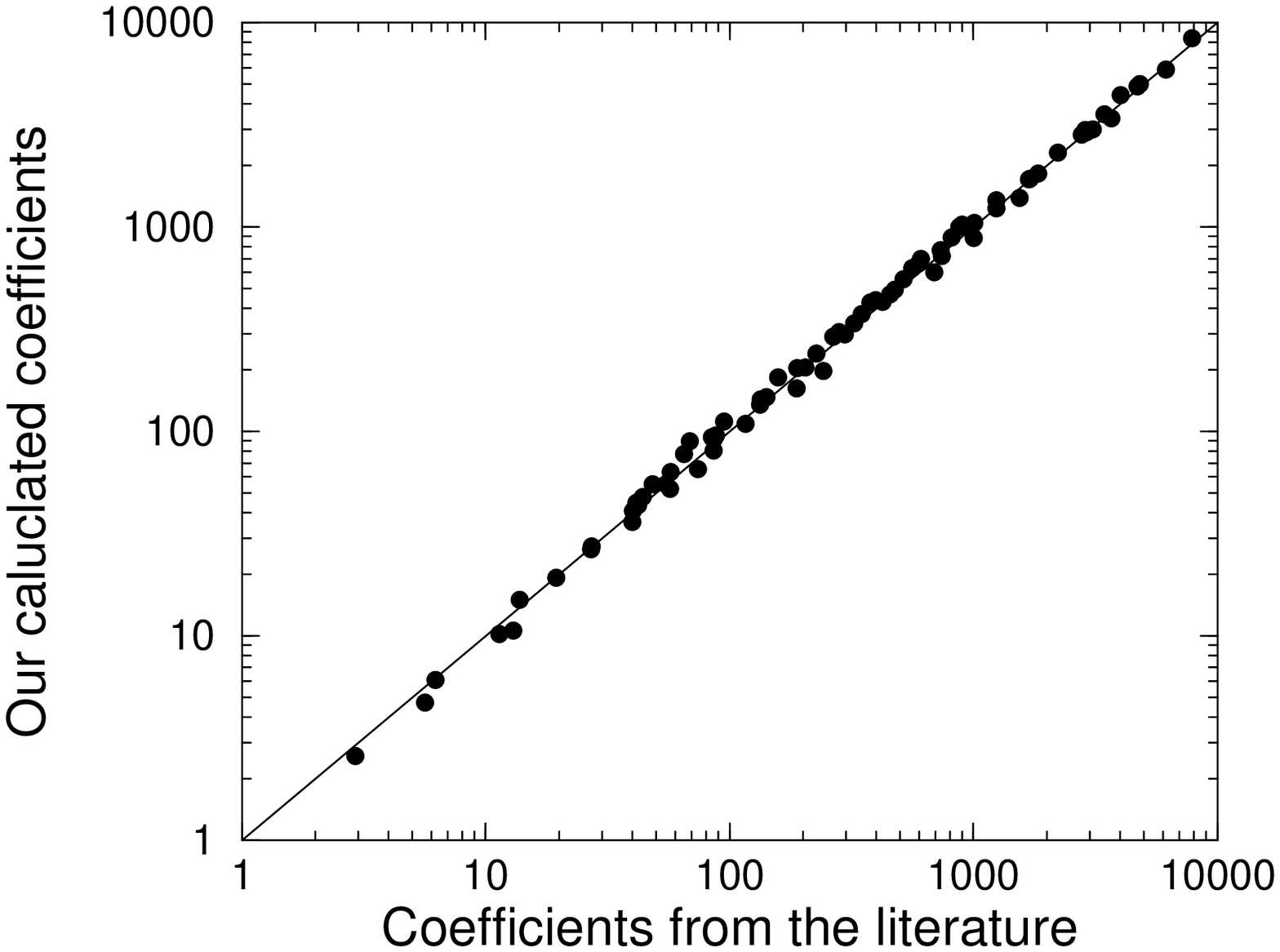,width=8.6cm}}
\label{wintergate}
\end{figure}

\begin{figure}
\caption{Our calculated $\alpha(iu)$ for He, compared with a more
accurate calculation.~\protect\cite{BiPi92}}
   \centerline{\psfig{figure=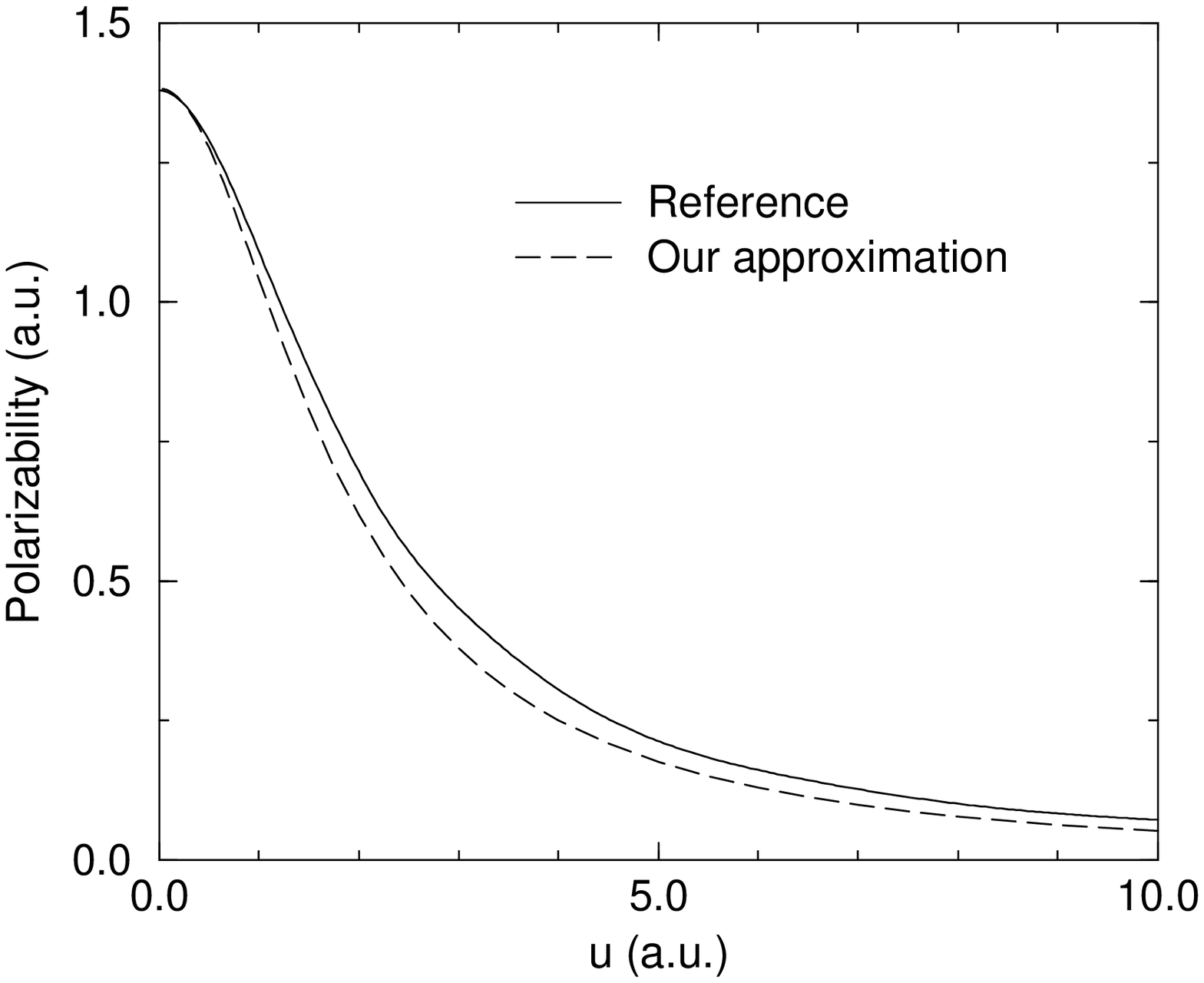,width=8.6cm}}
\label{Healpha}
\end{figure}

\begin{figure}
\caption{Our calculated $\alpha(iu)$ for Be, compared with a more
accurate calculation.~\protect\cite{KoHa91}}
   \centerline{\psfig{figure=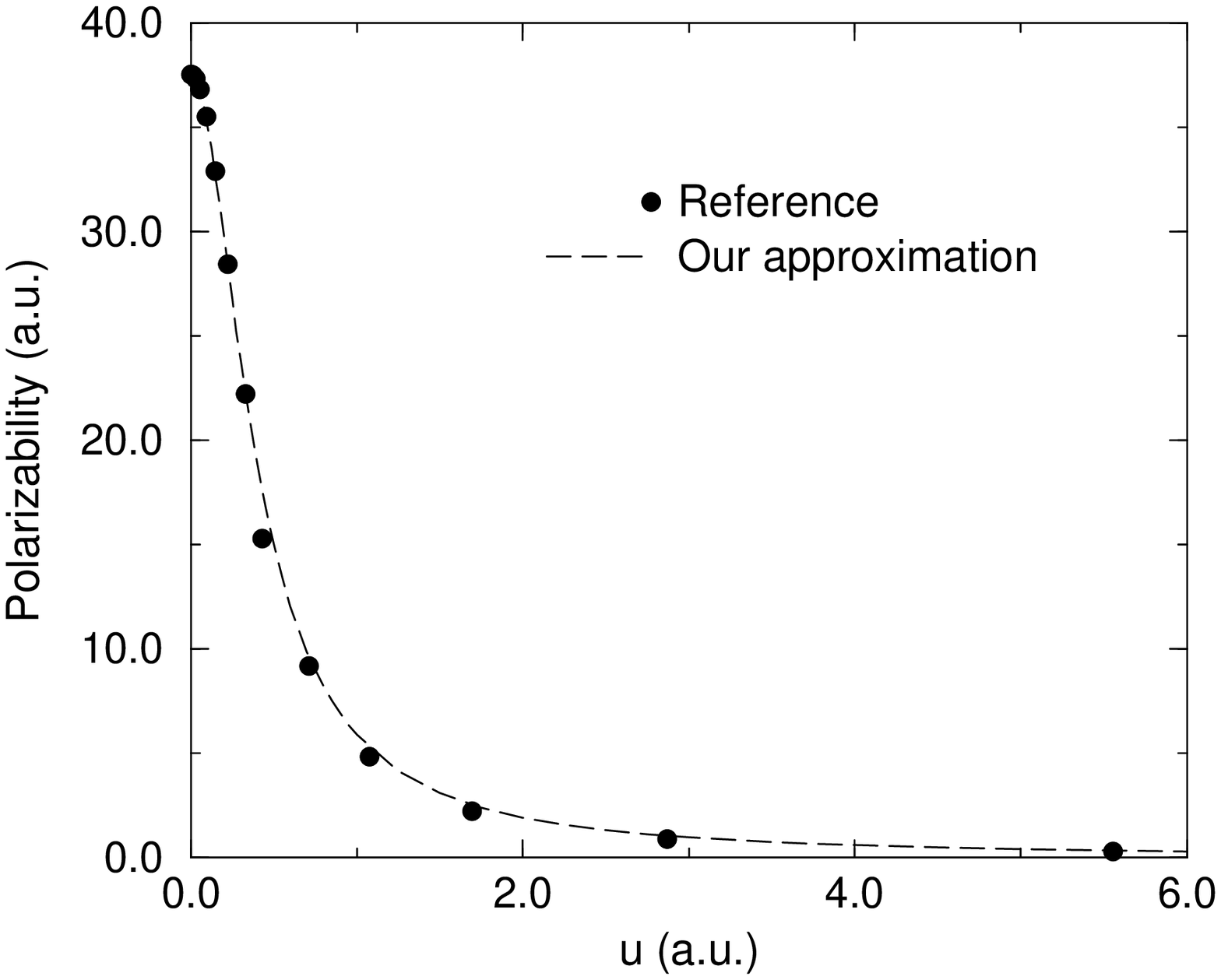,width=8.6cm}}
\label{Bealpha}
\end{figure}

The solution of Eq.~(\ref{eq:div}) is done with a finite element
method with an adaptive net.~\cite{Rylic} In this way we have a
general method for all geometries. 
To secure a reasonable numerical accuracy (5\%) at small frequencies, we here
represent $\alpha(iu)$ by the expression $a+b/(1+u^2/c^2)$, where
$a$, $b$ and $c$ are fixed by a smooth continuation of high-frequency
results.
In Table~\ref{sametab} calculated van der Waals coefficients for 
a number of pairs of identical atoms are given, together with the static
polarizability used when defining the cutoff. In
Figure~\ref{wintergate} calculated $C_6$ values both for identical
and mixed pairs of atoms are plotted against results from 
more accurate calculations.
The values compare very well with results from other calculations,
with the close agreement indicated by the narrow spread of the
points around the diagonal.
Especially the results for alkali and alkaline earth atoms are much improved
compared with our earlier calculations.~\cite{AnLaLu96} 
In Figures~\ref{Healpha} and \ref{Bealpha} the dynamic polarizability 
from our calculations are compared with reference calculations. 
Figure~\ref{Healpha} for He is a worst-case example, with
a 12\% error in $C_6$, while Figure~\ref{Bealpha} for Be is a best-case
example, with a $C_6$ right on the reference value.

Results for a few molecules are given in Table~\ref{moltab},
with results agreeing very well with literature values.
The largest molecule for which we have so far calculated the van der Waals
coefficient is fullerene, $C_{60}$.
Recently the dispersion energy between two fullerenes has been
computed from first principles in time-dependent DFT,
which gives the van der Waals coefficient $C_6=253$ kRy a$_0^6$.~\cite{PaPr97}
Earlier, simpler methods have been used to estimate the polarizability
and the van der Waals coefficient.
A summation of C--C interactions gives $C_6=200$ kRy
a$_0^6$,~\cite{GiLaDeLu94} and for a calculation of dipole modes
using a discrete dipole model the result is
$C_6=350$ kRy a$_0^6$.~\cite{GiLaDeLu94}
Using $\alpha(0)=570$ a$_0^3$ (experimental value from 
Ref.~\onlinecite{Aletal94})
we get $C_6=302$ kRy a$_0^6$, a result which lies in the same range as
those from the other calculations.

In Tables~\ref{sametab} and \ref{moltab} the characteristic
frequencies, $u_0$, corresponding to 
London's empirical formula~\cite{London30}
\begin{equation}
   C_6 = \alpha^A(0) \alpha^B(0) \frac{3 u_0^A u_0^B}
         {2(u_0^A + u_0^B)} \,,
\label{eq:London}
\end{equation}
where A and B denote the two fragments, are also given. 
This formula provides an easy way of estimating
isotropic van der Waals coefficients for mixed pairs of atoms
and molecules.~\cite{Mahan82,TaTo86}

\begin{figure}
\caption{Our calculated $\alpha_{zz}(iu)$ for $H_2$, compared with 
a more accurate calculation.~\protect\cite{BiPi92}}
   \centerline{\psfig{figure=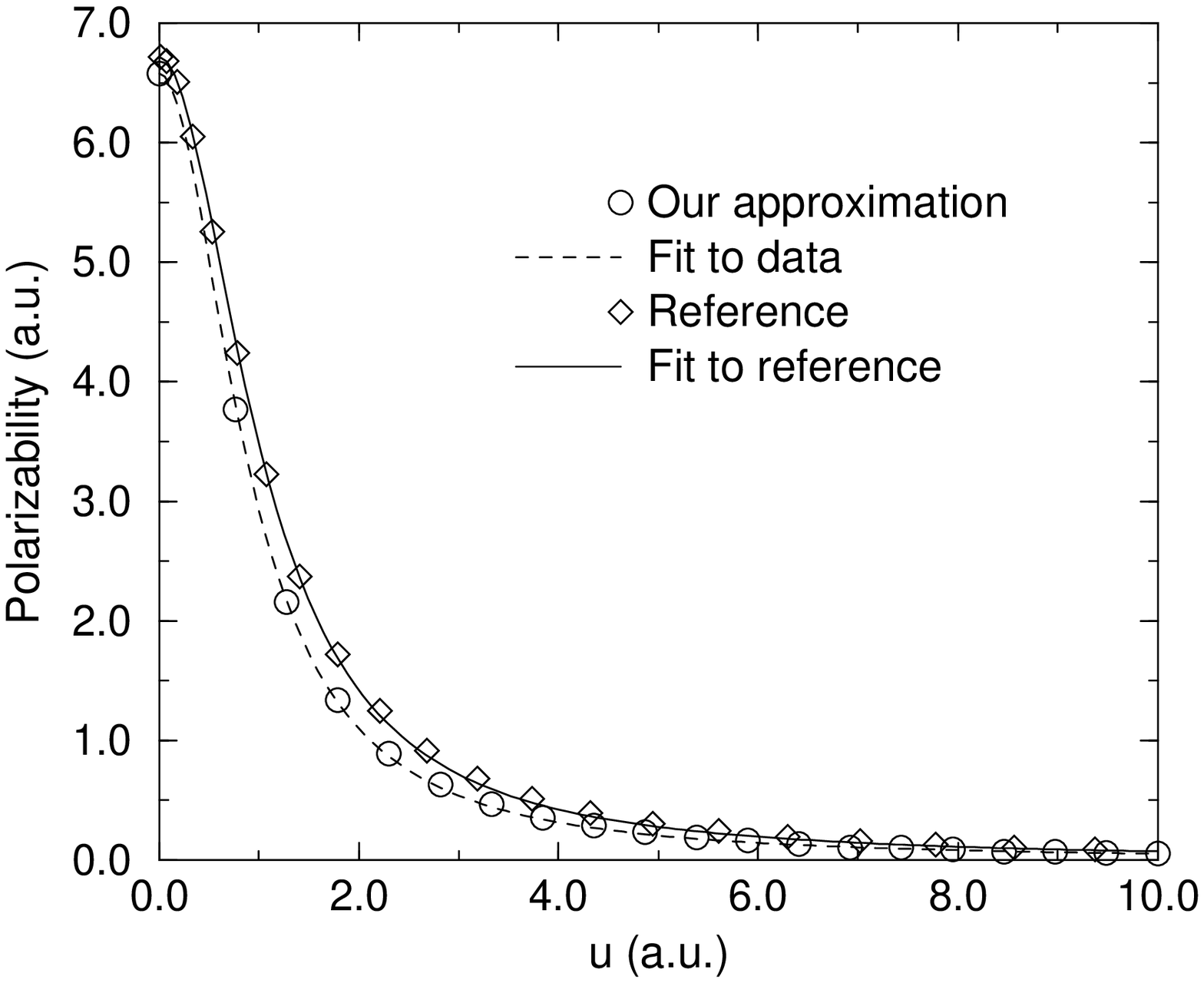,width=8.6cm}}
\label{H2zz}
\end{figure}

\begin{figure}
\caption{Our calculated $\alpha_{xx}(iu)$ for $H_2$, compared with 
a more accurate calculation.~\protect\cite{BiPi92}}
   \centerline{\psfig{figure=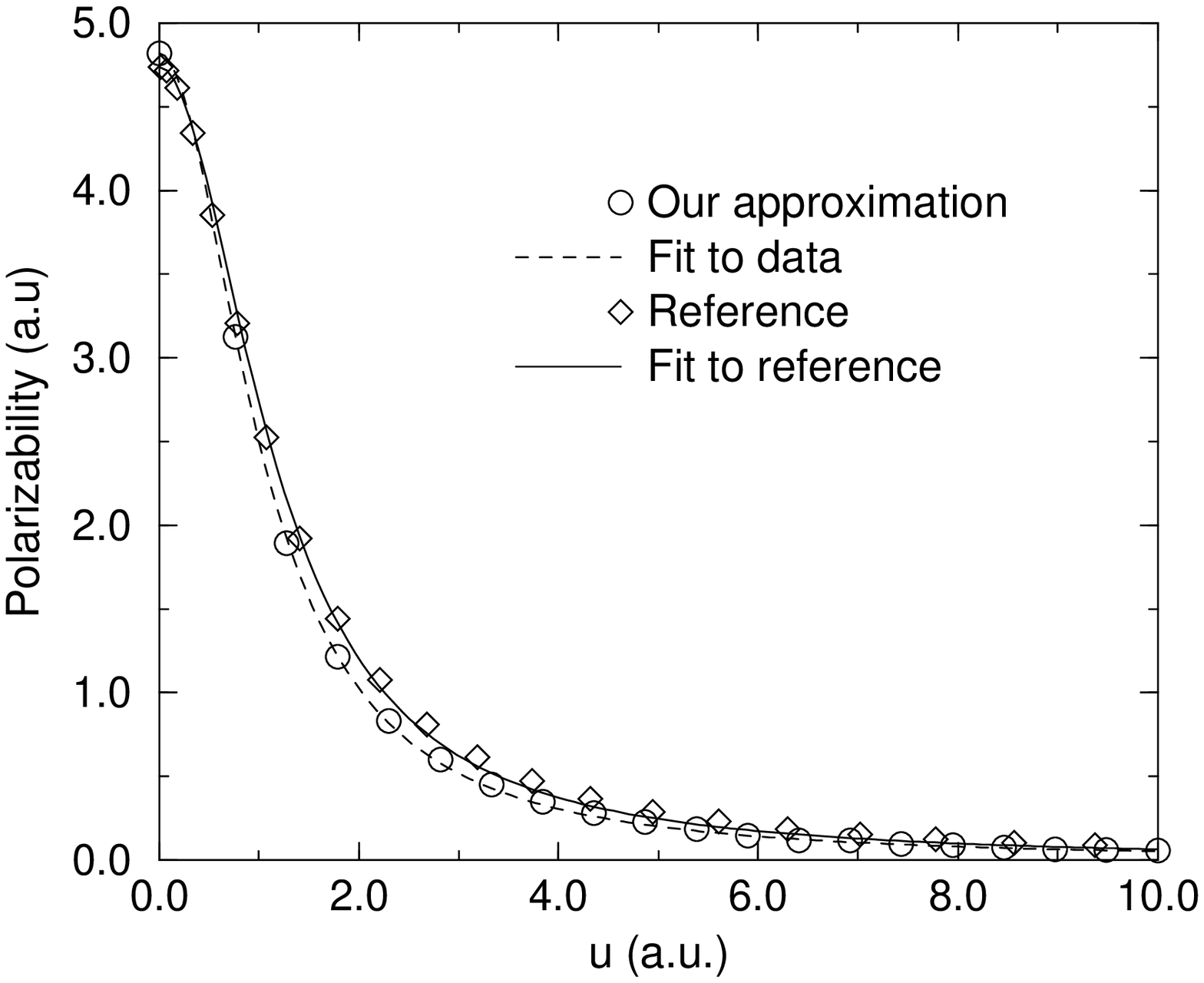,width=8.6cm}}
\label{H2xx}
\end{figure}

In Table~\ref{moltab} we have only given the isotropic dispersion coefficients
for the molecules, that is, we have used the averaged polarizability
$\overline{\alpha}(iu)$ in Eq.~(\ref{eq:C6}), where
\begin{equation}
   \overline{\alpha}(iu) = \frac{1}{3} [ \alpha_{xx}(iu) +
         \alpha_{yy}(iu) + \alpha_{zz}(iu) ] \,.
\label{alphabar}
\end{equation}
It is easy, however, to calculate the anisotropic corrections,
in addition.
A simple example is for two interacting identical linear molecules, where the
anisotropic coefficients $C_6^{\prime}$ and $C_6^{\prime \prime}$ control
the orientation-dependent part of the long-range interaction according
to~\cite{BiPi93}
\begin{eqnarray}
 & &  E_{\rm vdW}(R, \theta_A, \theta_B, \phi_A, \phi_B) = \nonumber \\
 & & - \left[ \rule{0cm}{0.5cm} C_6 + C_6^\prime\left\{P_2(\cos\theta_A)+
      P_2(\cos\theta_B)\right\}+ \frac{4 \pi}{5} C_6^{\prime \prime} \times \right.
      \nonumber \\
 & & \left. \sum_{m=-2}^{2}(3-|m|)
          Y_2^m(\theta_A, \phi_A) Y_2^{-m}(\theta_B,\phi_B) \right] R^{-6}\,,
\end{eqnarray}
where $\theta_A$ ($\theta_B$) is the angle between the vector ${\bf R}$
from the center of molecule $A$ to the center of $B$ and the axis
of molecule A (B).
The other angle $\phi_A$ ($\phi_B$) describes the rotation of molecule A (B)
about $R$.
With
\begin{equation}
  \Delta\alpha(iu) = \alpha_{zz}(iu)-\alpha_{xx}(iu)
\label{deltaalpha}
\end{equation}
these coefficients can be written~\cite{BiPi93}
\begin{equation}
   C_6^\prime = \frac{1}{\pi} \int_0^\infty du \, \overline{\alpha}_1(iu)
                 \Delta\alpha_2(iu)
\label{C6prime}
\end{equation}
and
\begin{equation}
   C_6^{\prime \prime} = \frac{1}{3 \pi} \int_0^\infty du \, 
                          \Delta\alpha_1(iu) \Delta\alpha_2(iu) \,.
\label{C6bis}
\end{equation}
Calculating the anisotropic coefficients for $H_2$ we obtain
$C_6^\prime/C_6 = 0.08$ and $C_6^{\prime \prime}/C_6=0.007$.
Accurate values are $C_6^\prime/C_6 = 0.1$ and
$C_6^{\prime \prime}/C_6=0.01$.~\cite{BiPi93} The anisotropy is thus
underestimated slightly with the simple cutoff scheme described above.
In Figures~\ref{H2zz} and~\ref{H2xx} our calculated $\alpha_{zz}(iu)$
and $\alpha_{xx}(iu)$ for $H_2$ are compared with accurate results.

For an atom or molecule a distance $d$ outside a surface,
the asymptotic
van der Waals energy is given by~\cite{ZaKo76}
\begin{equation}
  E_{\rm vdW} = -\frac{C_3}{(d-Z_0)^3} \,,
\end{equation}
where the van der Waals coefficient is
\begin{equation}
  C_3 = \frac{1}{4 \pi} \int_0^\infty \! du \, \overline{\alpha}(iu)
              \frac{\epsilon_b(iu)-1}{\epsilon_b(iu)+1} \,,
\label{eq:C3}
\end{equation}
and with the van der Waals plane
\begin{equation}
        Z_0 = \frac{1}{4 \pi C_3} \int_0^{\infty} du\, \alpha(iu)
               \frac{\epsilon_b(iu) - 1}{\epsilon_b(iu) + 1} 
              \frac{\epsilon_b(iu)}{\epsilon_b(iu)+1}\,d(iu) \,.
\end{equation}
In these expressions, $\epsilon_b(iu)$ is the bulk dielectric function,
and $d(iu)$ is the
centroid of the induced surface charge caused by an electric field
oriented perpendicular to the surface and varying in time like $e^{ut}$.
Our earlier calculations~\cite{HuAnLuLa96,HuKi97} of $C_3$ and $Z_0$
have used the exact electrodynamics, Eq.~(\ref{eq:PfromE}), for the surface
but the approximate treatment, Eq.~(\ref{eq:PfromEappl}), for the 
atom and molecule. In this paper we use Eq.~(\ref{eq:PfromE}) also for
the latter.
In Table~\ref{jelliumtab} calculated values for $C_3$ and $Z_0$ are given
for He and H$_2$ outside jellium, showing a very good agreement with
other more elaborate calculations. 

Including the orientational dependence that results from
the anisotropy of the molecular polarizability, the energy 
for a homonuclear diatomic molecule is to first order given by~\cite{HaFe82}
\begin{equation}
  E_{\rm vdW}(\theta) = -\frac{1}{d^3} \left[C_3^{(0)}+C_3^{(2)}
                     P_2(\cos\theta) \right] \,,
\end{equation}
where $\theta$ is the angle between the molecule axis and the surface
normal.
$C_3^{(0)}$ is given by Eq.~(\ref{eq:C3}) and 
\begin{equation}
  C_3^{(2)} = \frac{1}{4 \pi} \int_0^\infty \! du \, \Delta\alpha(iu)
              \frac{\epsilon_b(iu)-1}{\epsilon_b(iu)+1} \,.
\label{eq:C32}
\end{equation}
In Table~\ref{jelliumtab} the ratio $C_3^{(2)}/C_3^{(0)}$ is given
for H$_2$ outside jellium. We find this ratio to be
around 0.05 in agreement with Ref.~\onlinecite{HaFe82}.

We have in this paper refined the electrodynamical treatment
of atoms and molecules within our previously presented density
functional framework, thereby unifying our approaches for objects of
different sizes. The calculated polarizabilities and 
van der Waals coefficients
are in good agreement with results in the literature.
This gives a possibility to easily
calculate these quantities for complex systems
with useful accuracy.

\acknowledgments
Work at Rutgers supported in part by NSF Grants No. DMR 94-07055
and DMR 97-08499. Financial support from the Swedish Natural Science
Research Council and the Swedish Foundation for Strategical Research through
Materials Consortium no. 9 is also acknowledged.

\begin{table}
\caption{Van der Waals coefficients $C_6$ for pairs of identical atoms
(Ry atomic units). The static polarizabilities used for defining the cutoff
are given in the second column (atomic units) and results from other
calculations in the fifth one. The third column gives the frequency
$u_0$ obtained from the London formula, Eq.~(\ref{eq:London}).}
\begin{tabular}{c c c c c}
   & $\alpha(0)$ & $u_0$ & $C_6$ & $C_6^{\mbox{\scriptsize ref}}$
\\ \hline
He-He & 1.38$^a$ & 1.81     & 2.58 & 2.92$^b$ \\
Ne-Ne & 2.67$^a$ & 2.80     & 15.0 & 13.8$^c$ \\
Ar-Ar & 11.1$^a$ & 1.56     & 143    & 134$^c$ \\
Kr-Kr & 16.7$^a$ & 1.37     & 291    & 266$^c$ \\
Xe-Xe & 27.3$^d$ & 1.15     & 663   & 597$^c$ \\ 
 & & & & \\
H-H   & 4.5$^b$ & 0.70      & 10.6 & 13$^b$ \\
Li-Li & 164$^e$ & 0.14      & 2830 & 2780$^f$ \\
Na-Na & 159$^e$ & 0.16      & 3000 & 3080$^f$ \\
K-K   & 293$^e$ & 0.13      & 8400 & 7890$^f$ \\
 & & & & \\
Be-Be & 37.5$^g$   & 0.41   & 429  & 425$^g$  \\
Mg-Mg & 70$^h$     & 0.34   & 1230 & 1240$^f$ \\
Ca-Ca & 154$^f$    & 0.25   & 4430 & 4010$^f$ \\
\end{tabular}
\label{sametab}
$^a$ Ref.~\onlinecite{Leonard74}, $^b$ Ref.~\onlinecite{BiPi93}, $^c$ Ref.~\onlinecite{StCe85},
$^d$ Ref.~\onlinecite{TaTo86}, $^e$ Ref.~\onlinecite{MoScMiBe74}, $^f$ Ref.~\onlinecite{MaKu79},
$^g$ Ref.~\onlinecite{KoHa91}, $^h$ Ref.~\onlinecite{Gollisch84} \\
\end{table}

\begin{table}
\caption{Van der Waals coefficients $C_6$ for pairs of identical molecules
(Ry atomic units). The static polarizabilities used for defining the cutoff
are given in the second column (atomic units) and 
values from the literature in the fifth one. The third column gives the
frequency
$u_0$ obtained from the London formula, Eq.~(\ref{eq:London}).}
\begin{tabular}{c c c c c}
 & $\alpha(0)$ & $u_0 $ & $C_6$ & $C_6^{\mbox{\scriptsize ref}}$ \\ \hline
H$_2$-H$_2$ & 5.41$^a$ & 0.98  & 21.5 & 24.1$^a$ \\
N$_2$-N$_2$ & 11.77$^b$ & 1.44 & 149 & 147$^b$ \\
CO-CO & 13.1$^c$ & 1.37 & 176 & 163$^c$ \\
HF-HF & 5.52$^d$ & 1.77 & 40.4 & 38$^e$   \\
H$_2$O-H$_2$O & 9.64$^f$ & 1.40 & 97.4 & 93$^f$  \\
$C_{60}$-$C_{60}$ & 570$^g$ & 1.24 & 302$\,k$ & 200$\,k^h$, 253$\,k^i$, 
350$\,k^h$ \\
\end{tabular}
\label{moltab}
$^a$ Ref.~\onlinecite{BiPi93}, $^b$ Ref.~\onlinecite{MeKu90}, $^c$ Ref.~\onlinecite{KuMe94},
$^d$ Ref.~\onlinecite{DiSa81},
$^e$ Ref.~\onlinecite{KuMe85}, $^f$ Ref.~\onlinecite{ZeMe77},
$^g$ Mean value of estimates from Ref.~\onlinecite{Aletal94},
$^h$ Ref.~\onlinecite{GiLaDeLu94}
$^i$ Ref.~\onlinecite{PaPr97}
\end{table}

\begin{table}
\caption{The van der Waals coefficient $C_3$ and the van der Waals plane
position $Z_0$ (Ry atomic units) for He and H$_2$ interacting with 
jellium. For H$_2$ also the ratio between the anisotropic coefficient
$C_3^{(2)}$ and $C_3^{(0)}$ is given.}
\begin{tabular}{c c c c c c c}
 & $r_s$ & $C_3$ & $C_3^{\rm ref}$ & $C_3^{(2)}/C_3^{(0)}$ & $Z_0$ & $Z_0^{\rm ref}$ \\ \hline
He & 2 & 0.10 & 0.10$^a$ & & 0.78 & 0.74$^b$ \\ 
   & 3 & 0.064 & 0.064$^a$ & & 0.62 & 0.64$^b$ \\
   & 4 & 0.045 & 0.045$^a$ & & 0.53 & 0.59$^b$ \\ 
H$_2$ & 2 & 0.31 & 0.32$^a$ & 0.040 & 0.91 & 0.85$^b$ \\
      & 3 & 0.22 & 0.22$^a$ & 0.044 & 0.70 & 0.71$^b$ \\
      & 4 & 0.16 & 0.16$^a$ & 0.046 & 0.59 & 0.64$^b$ \\
\end{tabular}
\label{jelliumtab}
$^a$ Ref.~\onlinecite{ZaKo76},$^b$ Ref.~\onlinecite{Liebsch86}
\end{table}

\newpage

\newpage

\begin{thebibliography}{10}

\bibitem{HoKo64}
P. Hohenberg and W. Kohn, Phys. Rev. {\bf 136},  B864  (1964).

\bibitem{KoSh65}
W. Kohn and L.~J. Sham, Phys. Rev. {\bf 140},  A1133  (1965).

\bibitem{GuLu76}
O. Gunnarsson and B.~I. Lundqvist, Phys. Rev. B {\bf 13},  4274  (1976).

\bibitem{LaMe81}
D.~C. Langreth and M.~J. Mehl, Phys. Rev. Lett. {\bf 47},  446  (1981).

\bibitem{Becke88}
A.~D. Becke, Phys. Rev. A {\bf 30},  3098  (1988).

\bibitem{Peetal92}
J.~P. Perdew, J.~A. Chevary, S.~H. Vosko,
K.~A. Jackson, M.~R. Pederson, D.~J. Singh, and
C. Fiolhais, Phys. Rev. B {\bf 46},  6671  (1992).

\bibitem{PeBuEr96}
J.~P. Perdew, K. Burke, and M. Ernzerhof, Phys. Rev. Lett. {\bf 78},  3865
  (1996).

\bibitem{LuAnShChLa95}
B.~I. Lundqvist, Y. Andersson, H. Shao, S. Chan,
and D.~C. Langreth, Int. J. Quantum. Chem {\bf 56},  247  (1995).

\bibitem{AnLaLu96}
Y. Andersson, D.~C. Langreth, and B.~I. Lundqvist, Phys. Rev. Lett. {\bf 76},
  102  (1996).

\bibitem{HuAnLuLa96}
E. Hult, Y. Andersson, B.~I. Lundqvist, and D.~C. Langreth, Phys. Rev. Lett.
  {\bf 77},  2029  (1996).

\bibitem{DoDi96}
J.~F. Dobson and B.~P. Dinte, Phys. Rev. Lett. {\bf 76},  1780  (1996).

\bibitem{KoMeMa97}
W. Kohn, Y. Meir, and D.~E. Makarov, Phys. Rev. Lett. {\bf 80},
1998,  (4153).

\bibitem{Australia}
Y. Andersson, E. Hult, H. Rydberg, P. Apell, B.~I. Lundqvist, and
D.~C. Langreth,  in {\em Electronic Density Functional Theory:
  Recent Progress and New Directions}, edited by J.~F. Dobson, G. Vignale, and
  M.~P. Das (Plenum, New York, 1997).

\bibitem{LaPe7577}
D.~C. Langreth and J.~P. Perdew, Solid State Commun. {\bf 17},  1425 
(1975); Phys. Rev. B {\bf 15},  2884  (1977).

\bibitem{Dobson94}
J.~F. Dobson,  in {\em Topics in Condensed Matter Physics}, edited by M.~P. Das
  (Nova, N. Y., 1994).

\bibitem{RaAs91}
K. Rapcewicz and N.~W. Ashcroft, Phys. Rev. B {\bf 44},  4032  (1991).

\bibitem{AnRy97}
Y. Andersson and H. Rydberg, Density functional for van der Waals complexes,
  submitted to Physica Scripta.

\bibitem{HuKi97}
E. Hult and A. Kiejna, Surf. Sci. {\bf 383},  88  (1997).

\bibitem{AnHuApLaLu98}
Y. Andersson, E. Hult, P. Apell, D.~C. Langreth, and B.~I. Lundqvist,
Solid State Commun. {\bf 106},  235  (1998).

\bibitem{MaSt62}
C. Mavroyannis and M.~J. Stephen, Mol. Phys {\bf 5},  629  (1962).

\bibitem{MaKe69}
H. Margenau and N.~R. Kestner, {\em Theory of Intermolecular Forces} (Pergamon
  Press, Oxford, 1969).

\bibitem{Rylic}
H. Rydberg, Correlation Density Functional beyond the Atomic Scale, 1998,
  licentiate thesis, Chalmers University of Technology.

\bibitem{PaPr97}
J.~M. Pacheco and J.~P. \mbox{Prates Ramalho}, Phys. Rev. Lett. {\bf 79},  3873
   (1997).

\bibitem{GiLaDeLu94}
C. Girard, P. Lambin, A. Dereux, and A. Lucas, Phys. Rev. B {\bf 49},  11425
  (1994).

\bibitem{Aletal94}
F. Alasia, R.~A. Broglia, H.~E. Roman, L. Serra,
G. Colo, and J.~M. Pacheco, J. Phys. B {\bf 27},  L643  (1994).

\bibitem{London30}
F. London, Z. F. Phys. Chem. B {\bf 11},  222  (1930).

\bibitem{Mahan82}
G.~D. Mahan, J. of Chem. Phys. {\bf 76},  493  (1982).

\bibitem{TaTo86}
K.~T. Tang and J.~P. Toennis, Z. Phys. D {\bf 1},  91  (1986).

\bibitem{BiPi93}
D.~M. Bishop and J. Pipin, Int. J. Quantum Chem. {\bf 45},  349  (1993).

\bibitem{ZaKo76}
E. Zaremba and W. Kohn, Phys. Rev. B {\bf 13},  2270  (1976).

\bibitem{HaFe82}
J. Harris and P.~J. Feibelman, Surf. Sci. {\bf 115},  L133  (1982).

\bibitem{StCe85}
J.~M. Standard and P.~R. Certain, J. Chem. Phys {\bf 83},  3002  (1985).

\bibitem{MaKu79}
F. Maeder and W. Kutzelnigg, Chem. Phys. {\bf 42},  95  (1979).

\bibitem{KoHa91}
H. Koch and R.~J. Harrison, J. Chem. Phys. {\bf 95},  7479  (1991).

\bibitem{TaNoCe76}
K.~T. Tang, J.~M. Norbeck, and P.~R. Certain, J. Chem. Phys {\bf 64},  3063
  (1976).

\bibitem{BiPi92}
D.~M. Bishop and J. Pipin, J. of Chem. Phys {\bf 97},  3375  (1992).

\bibitem{Leonard74}
P.~J. Leonard, At. Nucl. Data Tables {\bf 14},  21  (1974).

\bibitem{MoScMiBe74}
R.~W. Molof, H.~L. Schwartz, T.~M. Miller, and B. Bederson, Phys. Rev. A {\bf
  10},  1131  (1974).

\bibitem{Gollisch84}
H. Gollisch, J. Phys. B {\bf 17},  1463  (1984).

\bibitem{MeKu90}
W.~J. Meath and A. Kumar, Int. J. Quantum Chem. Symp. {\bf 24},  501  (1990).

\bibitem{KuMe94}
A. Kumar and W.~J. Meath, Chem. Phys. {\bf 189},  467  (1994).

\bibitem{DiSa81}
G.~H.~F. Diercksen and A.~J. Sadlej, J. Chem. Phys. {\bf 75},  1253  (1981).

\bibitem{KuMe85}
A. Kumar and W.~J. Meath, Mol. Phys. {\bf 54},  823  (1985).

\bibitem{ZeMe77}
G.~D. Zeiss and W.~J. Meath, Mol. Phys. {\bf 33},  1155  (1977).

\bibitem{Liebsch86}
A. Liebsch, Phys. Rev. B {\bf 33},  7249  (1986).

\end{thebibliography}
\end{document}